\documentstyle[12pt,amsbsy]{article}
\newcommand{\beq}{\begin{equation}}
\newcommand{\eeq}{\end{equation}}

\newcommand{\beqn}{\begin{eqnarray}}
\newcommand{\eeqn}{\end{eqnarray}}

\begin{document}

\begin{titlepage}

\vspace{1cm}

\begin{center}
{\Large \bf  Budker Institute of Nuclear Physics}
\end{center}

\vspace{0.5cm}

\begin{flushright}
BINP 98-17\\
April 1998
\end{flushright}

\vspace{1cm}

\begin{center}
{\large \bf On Quantization of Black Holes}
\end{center}

\begin{center}
I.B. Khriplovich\footnote{khriplovich@inp.nsk.su}
\end{center}
\begin{center}
Budker Institute of Nuclear Physics\\
630090 Novosibirsk, Russia
\end{center}

\bigskip

\begin{abstract}
A simple argument is presented in favour of the equidistant spectrum 
in semiclassical limit for the horizon area of a
black hole. The following quantization 
rules for the mass $M_N$ and horizon area $A_{Nj}$ are
proposed:
\[ M_N = m_p \,[N(N+1)]^{1/4};\;\;
A_{Nj} = 8\pi l_p^2 \left[\sqrt{N(N+1)} + \sqrt{N(N+1) - j(j+1)}\;\right].\]
Here both $N$ and $j$ are nonnegative integers or half-integers.
\end{abstract}

\vspace{7.5cm}

\end{titlepage}

The quantization of black holes was proposed long ago in the 
pioneering work \cite{be}. The idea was based on the intriguing
observation \cite{cr} that the horizon area $A$ of a nonextremal black 
hole behaves in a sense as an adiabatic invariant. This last fact makes
natural the assumption that the horizon area should be quantized. Once 
this hypothesis is accepted, the general structure of the quantization 
condition for large (generalized) quantum numbers $n$ gets obvious, up 
to an overall numerical constant $\alpha$ (our argument here somewhat 
differs from that of the original paper \cite{be}). The quantization 
condition should be 
\beq\label{q}
A_n = \alpha l_p^2 n.
\eeq
Indeed, the presence of the Planck length squared
\beq
l_p^2 = { G \hbar \over c^3}
\eeq
in formula (\ref{q}) is only natural. Then, for $A$ to be finite in a 
classical limit, the power of $n$ in expression (\ref{q}) should be the 
same as that of $\hbar$ in $l_p^2$. The validity of this our argument 
can be checked 
by looking at any expectation value nonvanishing in the classical limit 
in usual quantum mechanics.
 
From different points of view the black hole quantization was discussed
later in Refs. \cite{mu,ko}. However, although a lot of work has been 
done since on the subject (see recent review \cite{bek}), still one cannot
say that the problem is 
solved. In particular, there are various prescriptions for the
numerical constant $\alpha$ in formula (\ref{q}), for instance,
$\alpha = 4 \ln2$ \cite{mu,bek,bem}; $\alpha = 8\pi$ \cite{ka}.
For $\alpha = 4 \ln2$, the mass spectrum looks as follows:
\beq\label{n}
M_n=\,\sqrt{{\ln 2 \over 4\pi}}\, m_p \sqrt{n},
\eeq
and the distance between neighbouring levels is
\beq\label{d}
M_n - M_{n-1} =\,\sqrt{{\ln 2 \over \pi}}\,{m_p^2 \over 4 M_n}.
\eeq
The emission spectrum of a black hole becomes discrete with the transition
frequencies being multiples of (\ref{d}). Its envelope corresponds to the 
Hawking temperature
\begin{equation}\label{T_m}
T\,=\,\frac{m_p^2 c^2}{8\pi M},
\end{equation} 
with the usual maximum (for bosons) at $T_{max}= 2.82\,T$.

We will address here the problem starting from the expression for the
horizon area of the Kerr black hole, which can be written as 
\begin{equation}\label{ke}
A = 8\pi l_p^2\left[{M^2 \over m_p^2} + \sqrt{{M^4 \over m_p^4}
 - j(j+1)}\;\right].
\end{equation}
Here
\beq
m_p = \left({\hbar c \over G}\right)^{1/2}
\eeq
is the Planck mass. As to the total angular momentum ${\bf J}$ of the
black hole, according to the firmly established quantum-mechanical rule 
for any isolated system, it should be quantized: 
\beq\label{qj}
{\bf J}^2 =\,\hbar^2 j(j+1). 
\eeq 
As usual, $j$ here is a nonnegative integer or half-integer. 

In the case of an extreme Kerr hole, equation
(\ref{ke}) leads to the following quantization rule for its
mass $M_e$ \cite{ma}:
\beq\label{me}
M_e = m_p \,[j(j+1)]^{1/4}.
\eeq  
Although for the extreme black hole the horizon area does not behave as 
an adiabatic invariant, it is natural to look for the 
quantization rule generalizing (\ref{me}) to nonextremal situations. 
Rather obvious generalization is:
\beq\label{im}
M_N = m_p\,[N(N+1)]^{1/4};
\eeq  
\begin{equation}\label{ia}
A_{Nj} = 8\pi l_p^2 \left[\sqrt{N(N+1)} + \sqrt{N(N+1) - j(j+1)}\;\right].
\end{equation}
Here $N$ is a nonnegative integer or half-integer which is the
maximum value of $j$. Of course, the dependence of the area $A_{Nj}$ on 
two quantum numbers $N$ and $j$, but not on one,
in no way contradicts its adiabatic properties, which are so crucial for 
the whole problem. 

If one goes over from $N$ to integer quantum numbers $n=2N$, the 
quantization rule (\ref{im}) becomes for large $n$
\beq\label{in}
M_n=\,{1 \over \sqrt{2}}\, m_p \sqrt{n},
\eeq
which differs from the quantization rule (\ref{n}) by numerical factor
$\sqrt{2\pi / \ln 2} = 3.01$.

For a Schwarzschild black hole a close quantization rule 
\beq\label{kan}
M_n=\,{1 \over 2}\, m_p \sqrt{n}
\eeq
with integer $n$ was obtained previously \cite{ka}, starting with 
periodic boundary conditions in Euclidean time.

One cannot but notice a certain resemblance between our formula (\ref{ia}) 
and the quantization condition for a surface area 
\begin{equation}\label{aa}
A_a = 8\pi l_p^2 \sum_i \sqrt{2j_{1i}(j_{1i}+1) + 2j_{2i}(j_{2i}+1)
- j_{12i}(j_{12i}+1)},
\end{equation}
obtained in the loop approach to quantum gravity \cite{al}. In formula 
(\ref{aa}) $j_{1i},\;j_{2i}$  are nonnegative integers 
or half-integers; $j_{12i}$ run from $|j_{1i}-j_{2i}|$ to
$j_{1i}+j_{2i}$. For usual closed surfaces both $\sum_i j_{1i}$ and
$\sum_i j_{2i}$ are integers.

\bigskip
\bigskip

I appreciate numerous discussions with A.A. Pomeransky, in particular
he attracted my attention to Ref. \cite{al}. I am grateful also to J.D. 
Bekenstein for the advice to
publish this note. The work was supported by the Russian 
Foundation for Basic Research through Grant No. 95-02-04436-Á.

\end{document}